\documentclass[a4paper,11pt]{article}
\usepackage[T2A]{fontenc}
\usepackage[utf8]{inputenc}
\usepackage[english]{babel}
\usepackage{amsmath}
\usepackage{amsfonts}
\usepackage{amssymb}
\usepackage{wasysym}
\usepackage{amsthm}
\usepackage{tikz}
\usepackage{tikz-cd}
\usetikzlibrary{decorations.markings}
\usetikzlibrary{positioning}
\usepackage{array}
\usepackage{ytableau}
\usepackage{longtable}
\usepackage{empheq}
\usepackage{multicol}
\usepackage{mathrsfs}
\usepackage{amssymb}
\usepackage{diagbox}
\usepackage{pb-diagram}
\usepackage{setspace}
\usepackage[hidelinks,colorlinks=true,unicode]{hyperref}
\usepackage{csquotes}
\hypersetup{linkcolor=blue, citecolor=blue, filecolor=blue, urlcolor=blue}
\usepackage{url}
\usepackage[left=2.3cm,right=2.3cm,top=2.3cm,bottom=2.3cm,bindingoffset=0cm]{geometry}
\usepackage[natbib=true, style=numeric, sorting=none]{biblatex}
\begin{document}

\title{{\Large {\bf Are Maxwell knots integrable?}
\vspace{.2cm}}
\author{
{\bf A.Morozov $^{a,b,c}$}\thanks{morozov.itep@mail.ru},
{\bf N.Tselousov $^{a,c}$}\thanks{tselousov.ns@phystech.edu}} \date{ }
}
\maketitle

\vspace{-5cm}

\begin{center}
	\hfill ITEP-TH-20/20\\
	\hfill IITP-TH-15/20\\
	\hfill MIPT-TH-13/20
\end{center}

\vspace{2.3cm}

\begin{center}

$^a$ {\small {\it Institute for Theoretical and Experimental Physics, Moscow 117218, Russia}}\\
$^b$ {\small {\it Institute for Information Transmission Problems, Moscow 127994, Russia}}\\
$^c$ {\small {\it Moscow Institute of Physics and Technology, Dolgoprudny 141701, Russia }} \\

\end{center}

\vspace{1cm}
\begin{abstract}
We review properties of the null-field solutions of source-free Maxwell equations. We focus on the electric and magnetic field lines, especially on limit cycles, which actually can be knotted and/or linked at every given moment. We analyse the fact that the Poynting vector induces self-consistent time evolution of these lines and demonstrate that the Abelian link invariant is integral of motion. The same is expected to be true also for the non-Abelian invariants (like Jones and HOMFLY-PT polynomials or Vassiliev invariants),
and many integrals of motion can imply that the Poynting evolution is actually integrable. We also consider particular examples of the field lines for the particular family of finite energy source-free "knot" solutions, attempting to understand when the field lines are closed -- and can be discussed in terms of knots and links. Based on computer simulations we conjecture that Ranada's solution, where every pair of lines forms a Hopf link, is rather exceptional. In general, only particular lines (a set of measure zero) are limit cycles and represent closed lines forming knots/links, while all the rest are twisting around them and remain unclosed. Still, conservation laws of Poynting evolution and associated integrable structure should persist.
\end{abstract}

\section{Introduction}
 
Knots are playing an increasingly important role in mathematical physics, but in phenomenological physics they remain rather exotic. This is despite knots served as an inspiration to many great physicists, starting at least from Lord Kelvin \cite{Kelvin}. In fact the reason can be simple: we are not looking around with enough attention to notice unconventional things. From this perspective, it was a great breakthrough when in 1989 A.F.Ranada demonstrated \cite{Ranada,Ranada2} that Hopf link appears naturally in a solution to the ordinary Maxwell equations, i.e. is no less natural than the conventional plane wave solution. In this paper we are going to discuss some surprising properties of such knot-revealing solutions and mention some open problems -- without going into important but relatively sophisticated mathematics, like in \cite{FaddevNiemi} or \cite{Lechtenfeld:2017tif, Kumar:2020xjr}. We begin in s.\ref{forcelines} from consideration of the field lines of electric and magnetic fields. When they are closed, they can be knotted or linked. A given field line evolves in time -- this evolution appears to proceed along the Poynting vector field, what, however imposes a non-trivial and non-expected condition on the parametrization --  we discuss these issues in s.\ref{timeevolution} and s.\ref{peculiarity}. Moreover, closed and linked field lines remain linked, and this is guaranteed by the conservation laws provided by knot invariants -- which in this context serve as equations of motion and can even make the evolution of field lines integrable -- see s.\ref{knotinvs} and s.\ref{integrability}. We illustrate this general consideration by analysis of the Ranada's Hopf solution in s.\ref{Hopfsol} and by a brief discussion of more general solutions in s.\ref{othersols}. In Hopf solution all field lines are closed but unknotted, and every pair of field lines forms a Hopf link  with no non-trivial links of higher order. However, in other solutions generic field lines are not closed, only some are -- but they can form quite non-trivial knots and links. There is still a question, which knots and links can appear and how abundant they are -- see \cite{Lechtenfeld:2017tif} for a possible approach to this problem. We end with a brief summary in s.\ref{conclusion}.  

\section{Field lines at a given time moment \label{forcelines}}
In this paper we review properties of solutions of Maxwell equations:
\begin{align}
  \label{Maxwell equations}
  \begin{aligned}
     \frac{\partial \textbf{B}}{\partial t} + \nabla \times \textbf{E}  &= 0, \hspace{10mm} &
    \nabla \cdot \textbf{B} &=0;
    \\
    \frac{\partial \textbf{E}}{\partial t} - \nabla \times \textbf{B}  &=0, \hspace{10mm} &
    \nabla \cdot \textbf{E} &=0.
  \end{aligned}
\end{align}
In particular, we are interested in the topological structure of the \textit{field lines} so let us define the field line itself. To be more specific, let us concentrate only on the electric field lines, however our analysis can be carried out for the magnetic field lines analogously. Let $\textbf{E}$ be an electric field at a given moment of time and $\textbf{x}(s)$ -- a smooth curve in some parametrization $s$. Then $\textbf{x}(s)$ is called an electric field line if the following condition holds for any point on the curve:
\begin{equation}
    \label{field line definition}
    \frac{d \textbf{x}(s)}{ds} \ \upuparrows \ \textbf{E}(\textbf{x}(s))
\end{equation}
In other words, the tangent vector of the curve should be collinear to the electric field at each point on the curve. We note that the parametrization of the field line is ambiguously defined. To proceed with practical calculation, we will fix a parametrization that is suitable for our purposes. For example, one may consider a field line in different parametrizations:
\begin{equation}
    \frac{d \textbf{x}(s_1)}{d s_1} = \textbf{E}(\textbf{x}(s_1)) 
    \hspace{10mm}
    \frac{d \textbf{x}(s_2)}{d s_2} = \frac{\textbf{E}(\textbf{x}(s_2))}{|\textbf{E}(\textbf{x}(s_2))|}
    \hspace{10mm}
    \frac{d \textbf{x}(s_3)}{d s_3} = \frac{\textbf{E}(\textbf{x}(s_3))}{|\textbf{E}(\textbf{x}(s_3))|^2}
    \hspace{10mm}
    \ldots
\end{equation}
All these equations define the same field line. The second parametrization $s_2$ corresponds to the length of the field line, but we mainly use the third one $s_3$ in our analysis of the knot invarinats.
\\
\\
At a given moment of time there are different types of field lines. The force lines with endpoints are prohibited by the zero divergence condition. In other words, endpoints of the force lines are electric/magnetic charges, which are absent in our case. Generally, several types are available: closed loops of finite length and endless curves of infinite length. The curves with an infinite length may go to infinity or be located in the finite area. \\
Closed field lines are in fact limit cycles. It would be interesting to find sufficient conditions of closedness of the field lines. We discuss particular examples of field lines in s.\ref{Hopfsol}, s.\ref{othersols}.

\section{Time evolution of force lines \label{timeevolution}}

\begin{figure}[ht]
\centering
\setlength{\unitlength}{1.3cm}
\begin{picture}(10,4)
\thicklines
\qbezier(0,2)(4,5)(8,1.5)
\qbezier(0,1)(4,3)(8,0.5)
\put(8.2,0.6){{ \footnotesize a field line at}}
\put(8.2,0.3){{ \footnotesize the moment $t$}}
\put(8.2,1.6){{ \footnotesize a field line at}}
\put(8.2,1.3){{ \footnotesize the moment $t + dt$}}
\put(2.9,1.57){{$\textbf{x}_{1}$}}
\put(4.4,1.55){{$\textbf{x}_{2}$}}
\put(2.6,3.45){{$\textbf{x}_{1}^{\prime}$}}
\put(4.6,3.4){{$\textbf{x}_{2}^{\prime}$}}
\put(3,1.88){\vector(-0.1,0.85){0.166}}
\put(4.5,1.84){\vector(0.1,0.85){0.169}}
\put(1.7,2.4){{$\textbf{v}(\textbf{x}_{1}) dt$}}
\put(4.8,2.4){{$\textbf{v}(\textbf{x}_{2}) dt$}}
\end{picture}
\caption{\small The picture demonstrates the condition that the vector field \textbf{v} should obey to define self-consistent time evolution of the field lines. For any two points on a field line at the moment $t$ the ends of the vectors \textbf{v}$dt$ at the corresponding points lie on a field line that is defined at the moment $t + dt$.}
\label{field line time evolution pic}
\end{figure}
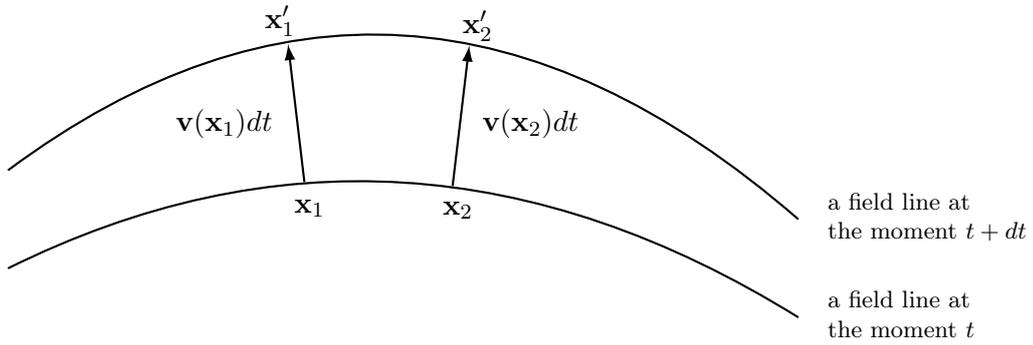

In our paper we consider only \textit{null-field} solutions of Maxwell equations. This means that at every moment of time and at every point in space the following conditions hold:
\begin{equation}
    \label{null field condition}
     \textbf{E} \cdot \textbf{B} = 0, \hspace{10mm} \textbf{E}^2 = \textbf{B}^2.
\end{equation}
Generally, Maxwell equations do not preserve these conditions, which makes them non-trivial. However, such solutions exist and a procedure of constructing them was described in detail in s.7 of \cite{Kumar:2020xjr}. The simplest solution of this kind is the Ranada's Hopf solution and we discuss it in s.\ref{Hopfsol}. 
\\
\\
We mainly focus on the following property of null-field solutions. It turns out that one can define self-consistent time evolution of the field lines \cite{NEWCOMB1958347, Irvine_2010}.  Generally, at different moments of time the space is filled with different field lines, due to the electric field evolution according to Maxwell equations. To give a prescription of how a field line deforms and moves in space, one needs to define a velocity vector $\textbf{v}$ for each point of a field line. This vector $\mathbf{v}$ is normal to the field line at each point. This structure will match the field lines at different moments of time. 
\\
\\
However, to define the time evolution of the field lines, a velocity vector field should obey a very special, yet natural condition. The condition requires that  any two points of a field line become points of another field line in the further moment of time when moved along the velocity vector field (see Fig.\ref{field line time evolution pic}). In other words, points of a field lines can not become points of different field lines. Alternatively, one could require that the following curve 
\begin{equation}
    \textbf{x}^{\prime}(s) = \textbf{x}(s) + \textbf{v}(\textbf{x}(s)) dt
\end{equation}
is a field line that is defined at the moment $t + dt$.
\\
In case of null-field solutions, the velocity vector $\textbf{v}$ can be chosen as the \textit{normalized} Poynting vector:
\begin{equation}
    \label{Poynting vector}
    \textbf{p} = \frac{\textbf{E} \times \textbf{B}}{|\textbf{E}| |\textbf{B}|}
\end{equation}
\subsection{Poynting time evolution \label{Poyting time evolution}}
In this section we demonstrate that the Poyting vector in a null-field solution induces self-consistent time evolution of the force lines. To do this, we explicitly check the self-consistency of the condition from the previous section.
\\
\\
We consider an electric field line $\textbf{x}(s)$  of a null-field solution at the moment $t$. In this section we fix the parametrization $s$ by the following equation:
\begin{equation}
    \frac{d \textbf{x}(s)}{ds} = \textbf{E}(\textbf{x}(s), t)
\end{equation}
Then we consider an auxiliary line $\textbf{x}^{\prime}(s)$ that is obtained from $\textbf{x}(s)$ by a slight shift along the Poynting vector $\textbf{p}$:
\begin{equation}
    \textbf{x}^{\prime}(s) = \textbf{x}(s) + \textbf{p}(\textbf{x}(s),t) dt
\end{equation}
Note that we shift along the Poynting vector by $dt$. The main claim is that $\textbf{x}^{\prime}(s)$ coincides as geometrical objects with some field line at the moment $t + dt$. Namely, a tangent vector of the line $\textbf{x}^{\prime}(s)$ is \textit{collinear} to the electric field at the moment $t + dt$, it means that they define the same field line:
\begin{equation}
\frac{d \textbf{x}^{\prime}(s)}{ds} \ \upuparrows \ \textbf{E}(\textbf{x}^{\prime}(s), t + dt)
\end{equation}
We should note that this collinearity holds only in the first order in $dt$. To explicitly see this, we compute the corresponding cross product:
\begin{equation}
\label{collinearity}
\begin{split}
    \frac{d \textbf{x}^{\prime}(s)}{ds} \times \textbf{E}(\textbf{x}^{\prime}(s), t + dt) & = 
    \Biggl( 
    \textbf{E} + 
    (\textbf{E} \cdot \nabla) \textbf{p} \, dt
    \Biggr) 
    \times
    \Biggl(
    \textbf{E} + 
    (\textbf{p} \cdot \nabla) \textbf{E} \, dt + \frac{\partial \textbf{E}}{\partial t} dt + o(dt) 
    \Biggr) = \\
    & = dt \, \textbf{E} \times \Biggl(  (\textbf{p} \cdot \nabla) \textbf{E} - (\textbf{E} \cdot \nabla) \textbf{p} + \frac{\partial \textbf{E}}{\partial t}  \Biggr) + o(dt) = o(dt)
\end{split}
\end{equation}
To simplify the formulas, we omit the explicit coordinate and time dependence of the fields meaning $\textbf{E} = \textbf{E}(\textbf{x}(s), t)$ and $\textbf{p} = \textbf{p}(\textbf{x}(s), t)$. We compute the expression in the brackets using the following identity for null-field solutions:
\begin{equation}
    \label{EP - PE}
     \text{E}_{\beta} \text{p}_{\alpha} 
        - \text{E}_{\alpha} \text{p}_{\beta} 
        = \epsilon_{\alpha \beta \gamma} \text{B}_{\gamma} 
\end{equation}
Taking the derivative of (\ref{EP - PE}) and using the Maxwell equations, we obtain the expression of the form:
\begin{equation}
\label{the evolution with divP}
    (\textbf{p} \cdot \nabla) \textbf{E} - (\textbf{E} \cdot \nabla) \textbf{p} + \frac{\partial \textbf{E}}{\partial t} = - (\nabla \cdot \textbf{p}) \textbf{E}
\end{equation}
The r.h.s is collinear to the electric field $\textbf{E}$ that ensures (\ref{collinearity}).
Finally, we conclude that the Poynting vector defines the self-consistent time evolution of the field line. Therefore, we can think of a field line as a strand where each point moves with the velocity vector $\textbf{p}$. From now we understand the time dependence of the field lines $\textbf{x}(s,t)$ as it is induced by the Poynting vector:
\begin{equation}
\label{Poynting evol}
    \frac{d \textbf{x}(s,t)}{dt} = \textbf{p}(\textbf{x}(s,t),t)
\end{equation}
It is evident now why we consider the normalized Poyting vector. If one chooses another normalization for the Poynting vector it will spoil (\ref{the evolution with divP}).
\subsection{Examples}
For demonstrative purposes let us discuss plane wave null-field solutions of Maxwell equations to analyse the structure of the field lines and its time evolution induced by the Poynting vector.
\begin{itemize}
    \item \textbf{Plane wave with linear polarization}\\
    The plane wave with linear polarization has the form:
    \begin{equation}
    \textbf{E} = 
    \left(\begin{array}{c}
        E \cos (z - t) \\
        0 \\
        0
    \end{array}\right) 
    \hspace{10mm}
    \textbf{B} = 
    \left(\begin{array}{c}
        0 \\
        E \cos (z - t) \\
        0
    \end{array}\right)
    \hspace{10mm}
    \textbf{p} = 
    \left(\begin{array}{c}
    0\\
    0\\
    1
    \end{array}\right)
    \end{equation}
Note that at a given moment of time the the electric field changes along the $z$-axis. However, all the electric field lines are straight lines parallel to the $x$-axis.  The field lines  fill the whole space except "singular" $xy$-planes with coordinates $z = t + \pi/2 + \pi n, n \in \mathbb{Z}$. We can write an explicit formula for the electric field line that goes through the point $(y_0, z_0)$ in the $yz$-plane at the moment $t_0$:
    \begin{equation}
    \textbf{x}(s,t) = 
    \left(\begin{array}{c}
    s \, E \cos (z_0 - t_0) \\
    y_0 \\
    z_0 + t - t_0
\end{array}\right)
\end{equation}
At each moment of time the tangent vector of the field line is the electric field:
\begin{align}
    \frac{d \textbf{x}(s,t)}{ds} = \textbf{E}(\textbf{x}(s,t),t) \hspace{10mm}
    \Leftarrow \hspace{10mm}
    \left(\begin{array}{c}
    E \cos (z_0 - t_0) \\
    0 \\
    0
\end{array}\right) =
\left(\begin{array}{c}
    E \cos (z_0 + t - t_0 - t) \\
    0 \\
    0
\end{array}\right) 
\end{align}
We note that the tangent vector of the particular field line does not change over time. This is due to the fact $ \nabla \cdot \textbf{p} = 0$ and we discuss this in s.\ref{peculiarity}.
The velocity of the points of the field line is the Poynting vector:
\begin{align}
    \frac{d \textbf{x}(s,t)}{dt} = \textbf{p}(\textbf{x}(s,t),t) \hspace{10mm}
    \Leftarrow \hspace{10mm}
    \left(\begin{array}{c}
    0 \\
    0 \\
    1
\end{array}\right) =
\left(\begin{array}{c}
    0 \\
    0 \\
    1
\end{array}\right)
\end{align}
The structure of the field lines flows along the $z$-axis over time as a whole and do not change. This is evident from the coordinate independence of the Poynting vector.
\item \textbf{Plane wave with elliptic polarization} \\
This solution has the form:
\begin{equation}
    \textbf{E} = \left(\begin{array}{c}
    E_x \cos (z - t) \\
    E_y \sin (z - t) \\
    0
\end{array}\right) 
\hspace{10mm}
\textbf{B} = \left(\begin{array}{c}
    -E_y \sin (z - t) \\
    E_x \cos (z - t) \\
    0
\end{array}\right)
\hspace{10mm}
\textbf{p} = \left(\begin{array}{c}
    0\\
    0\\
    1
\end{array}\right)
\end{equation}
This solution of Maxwell equations has a slightly more complicated structure of the field lines, however the time evolution is the same as in the previous example -- the field lines move along the $z$-axis, while the structure of the field lines remains the same. The explicit formula for the field line that goes through the point $(x_0, y_0, z_0)$ at the moment $t_0$:
\begin{equation}
    \textbf{x}(s,t) = \left(\begin{array}{c}
    x_0 + s \ E_x \cos (z_0 - t_0) \\
    y_0 + s \ E_y \sin (z_0 - t_0) \\
    z_0 + t - t_0
\end{array}\right)
\end{equation}
The field lines are normal to the $z$-axis and all field lines on a particular $xy$-plane are parallel. However, the direction of the field lines at a fixed moment of time is continuously changing along the $z$-axis. Namely, the direction rotates clockwise or anticlockwise along the $z$-axis at a fixed moment of time depending on the mutual sign of $E_x, E_y$.\\
At each moment of time the tangent vector is the electric field and the velocity is the Poynting vector:
\begin{align}
    \frac{d \textbf{x}(s,t)}{ds} = \textbf{E}(\textbf{x}(s,t),t) \hspace{10mm}
    \Leftarrow \hspace{10mm}
    \left(\begin{array}{c}
    E_x \cos (z_0 - t_0) \\
    E_y \sin (z_0 - t_0) \\
    0
\end{array}\right) =
\left(\begin{array}{c}
    E_x \cos (z_0 + t - t_0 - t) \\
    E_y \sin (z_0 + t - t_0 - t)\\
    0
\end{array}\right) 
\end{align}
\begin{align}
    \frac{d \textbf{x}(s,t)}{dt} = \textbf{p}(\textbf{x}(s,t),t) \hspace{10mm}
    \Leftarrow \hspace{10mm}
    \left(\begin{array}{c}
    0 \\
    0 \\
    1
\end{array}\right) =
\left(\begin{array}{c}
    0 \\
    0 \\
    1
\end{array}\right)
\end{align}

\item \textbf{Electromagnetic knot} \\
We consider a member of the family of finite energy source-free knot solutions of Maxwell equations \cite{Lechtenfeld:2017tif, Kumar:2020xjr}. We do not provide the explicit form of the electric and magnetic fields, because its is too complicated. The crucial point is that we use the null-field solution.
Instead of formulas we provide a computer simulation. We consider a closed field lines - limit cycles, that represent the trefoil knots. The little yellow knot becomes the big red one under the time evolution (see Fig.\ref{trefoil evolution}). The green curves are the trajectories of points and obey equation (\ref{Poynting evol}).\\
We argue that this picture represents general features of the Poynting evolution:
\begin{enumerate}
    \item The Poynting evolution is well defined. The green trajectories ensure the self-consistency condition.
    \item Limit cycles remain limit cycles. It is a strong argument in support of the Poynting evolution, because the time evolution of the limit cycles is well defined without any additional structure. Indeed, generally limit cycles are distinguished field lines, because they are closed. The Poynting time evolution \textit{coincides} with the evolution of the limit cycles.
    \item The topological structure of the limit cycles remains the same. Indeed, in both cases (yellow and red) the knot is the trefoil knot $3_1$.
\end{enumerate}
\begin{figure}[h!]
    \centering
    \includegraphics[scale=0.255]{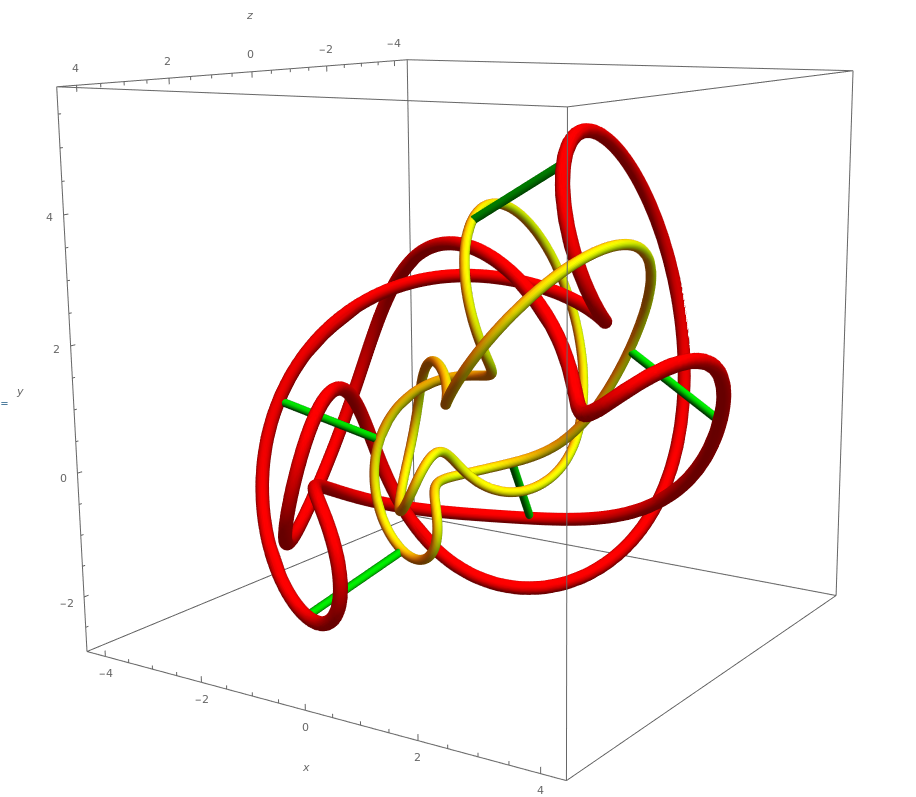}
    \includegraphics[scale=0.255]{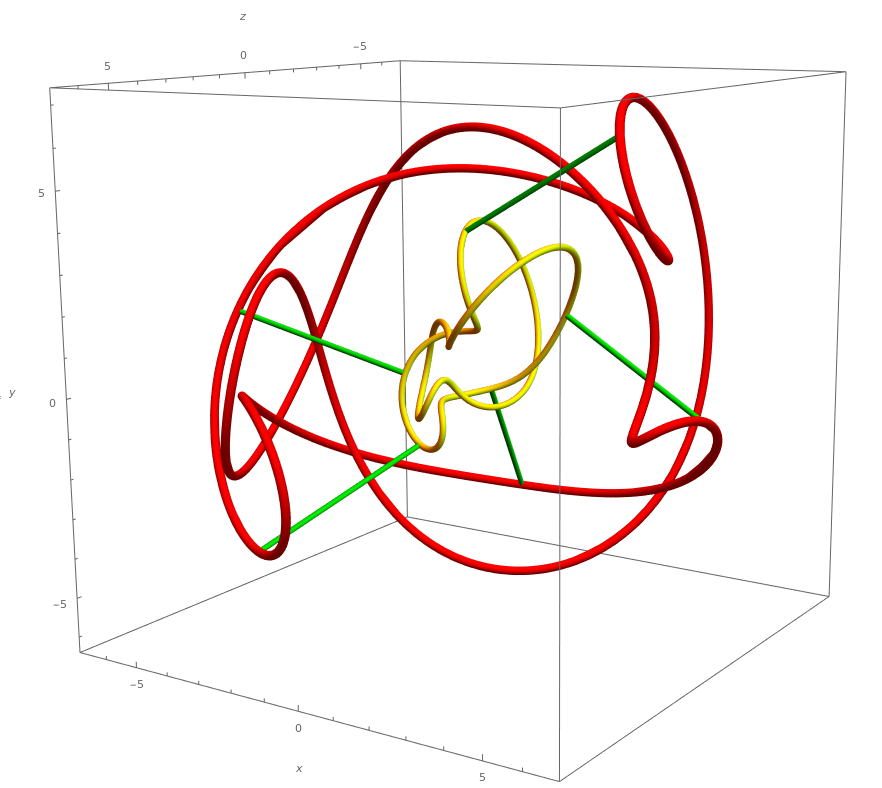}
    \caption{\footnotesize The little yellow knot becomes the big red one under time evolution. There are pictures of two successive moments of time. The green curves are the trajectories. The topological structure of knot does not change under the time evolution.}
    \label{trefoil evolution}
\end{figure}
\end{itemize}

\subsection{Peculiarities of Poynting evolution \label{peculiarity}}
\begin{figure}[ht]
\centering
\setlength{\unitlength}{1.3cm}
\begin{picture}(8,3)
\thicklines
\put(2,0){{$\textbf{x}$}}
\put(2.1,0.3){\vector(0,1){2}}
\put(6.1,0.3){\vector(0,1){2}}
\put(2.1,0.3){\vector(1,0){4}}
\put(2.1,2.3){\vector(1,0){4}}
\put(0.9,1.2){{$\textbf{p}(\textbf{x}, t) dt$}}
\put(6.3,1.2){{$\textbf{p}(\textbf{x} + \textbf{E}(\textbf{x},t) ds) dt$}}
\put(3.6,0.5){{$\textbf{E}(\textbf{x}, t) ds$}}
\put(1.6,2.5){{$\textbf{E}(\textbf{x} + \textbf{p}(\textbf{x}, t) dt, t + dt) ds (1 + (\nabla \cdot \textbf{p}) \, dt)$}}
\end{picture}
\caption{\footnotesize{Graphical interpretation of the self-consistency condition. The tangent vector of the auxiliary line is collinear to the electric field.}}
\label{parallelogram 1}
\end{figure}
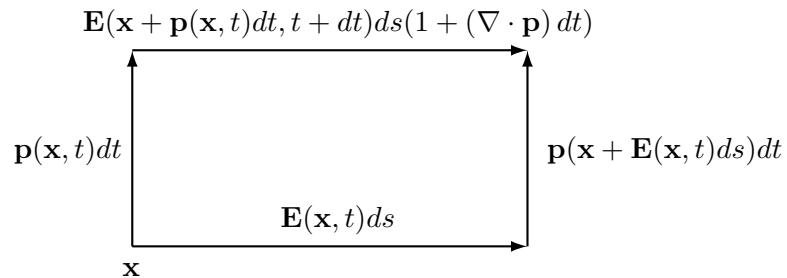

\begin{figure}[ht]
\centering
\setlength{\unitlength}{1.3cm}
\begin{picture}(8,3)
\thicklines
\put(2,0){{$\textbf{x}$}}
\put(2.1,0.3){\vector(0,1){2}}
\put(6.1,0.3){\vector(0,1){2}}
\put(2.1,0.3){\vector(1,0){4}}
\put(2.1,2.3){\vector(1,0){4}}
\put(0.9,1.2){{$\textbf{p}(\textbf{x}, t) dt$}}
\put(6.3,1.2){{$\textbf{p}(\textbf{x} + \textbf{e}(\textbf{x},t) ds) dt$}}
\put(3.6,0.5){{$\textbf{e}(\textbf{x}, t) ds$}}
\put(2.6,2.5){{$\textbf{e}(\textbf{x} + \textbf{p}(\textbf{x}, t) dt, t + dt) ds$}}
\end{picture}
\caption{\footnotesize{Graphical interpretation of the self-consistency condition for rescaled fields. The tangent vector of the auxiliary line coincides with the electric field.}}
\label{parallelogram 2}
\end{figure}
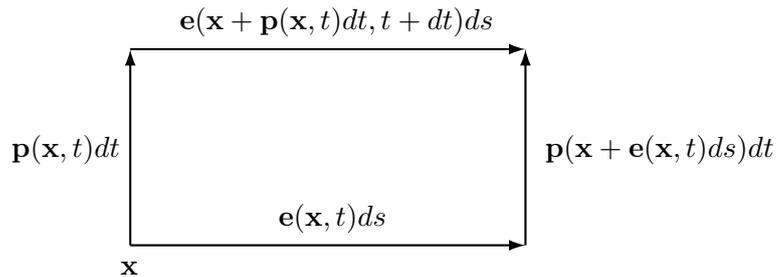

In this section we derive the parametrization of the field lines that is suitable for our further analysis of the link invariants. Also we argue that in this parametrization the commutativity of flows along $s$ and $t$ is evident. 
\\
\\
As has been shown in the section \ref{Poyting time evolution}, the tangent vector of the auxiliary line
\begin{equation}
    \textbf{x}^{\prime}(s) = \textbf{x}(s) + \textbf{p}(\textbf{x}(s),t) dt
\end{equation}
is not equal to the electric field at the moment $t + dt$ but only collinear. Graphically it can be shown in the following parallelogram of vectors \ref{parallelogram 1}.
This parallelogram is the consequence of the formula (\ref{the evolution with divP}). As can be seen, the upper vector is a slightly rescaled electric field. This fact can be thought of as the change in the parametrization:
\begin{equation}
    ds^{\prime} = ds (1 + (\nabla \cdot \textbf{p}) dt)
\end{equation}
We are free to multiply the electric field by a function $f(\textbf{E}^2)$  and the corresponding field lines will remain the same. However, the function $f$ can be chosen in such a way that the parametrization does not change:
\begin{equation}
    ds^{\prime} = ds
\end{equation}
In other words, the tangent vector of the auxiliary line (the upper vector in the parallelogram) will be equal to the vector $\textbf{E} f(\textbf{E}^2)$ on the shifted line at the moment $t + dt$. Namely:
\begin{align}
    \frac{d \textbf{x}(s)}{ds} &= \textbf{E}(\textbf{x},t) \ f(\textbf{E}^2(\textbf{x},t)) \label{cond 1} \\
    \frac{d \textbf{x}^{\prime}(s)}{ds} &= \textbf{E}(\textbf{x}^{\prime},t + dt) \ f(\textbf{E}^2(\textbf{x}^{\prime},t + dt)) \label{cond 2}
\end{align}
From the last equation in the first order in $dt$ we obtain the following constraint on the function $f$:
\begin{equation}
    (\nabla \cdot \textbf{p}) 
    \left(
    f + \textbf{E}^2 \frac{\partial f}{\partial \textbf{E}^2} 
    \right) = 0
\end{equation}
Note that if $(\nabla \cdot \textbf{p}) = 0$, the equation (\ref{cond 2}) will be satisfied automatically with any function $f$. Otherwise, if $(\nabla \cdot \textbf{p}) \not = 0$, the function is determined uniquely:
\begin{equation}
    f(\textbf{E}^2) = \frac{1}{\textbf{E}^2}
\end{equation}
To simplify the formulas, let us introduce the rescaled fields:
\begin{align}
    \textbf{e} &:= \textbf{E} / \textbf{E}^2 \label{e = E/E^2} 
\end{align}
Note that the rescaling of the fields does not spoil the null field conditions (\ref{null field condition}). In these new notations both the analog of the formula (\ref{the evolution with divP}) and the parallelogram \ref{parallelogram 2} look simpler and do not contain the term with $(\nabla \cdot \textbf{p})$:
\begin{equation}
\label{the evolution without divP}
     (\textbf{p} \cdot \nabla) \textbf{e}  + \frac{\partial \textbf{e}}{\partial t} = (\textbf{e} \cdot \nabla) \textbf{p}
\end{equation}
This formula is crucial for our analysis of the topological structure of the field lines. The rescaling of the electric fields does not change the structure of the field lines, however the formulas and calculations in the new notations are much simpler.
\\
As was discussed in the previous section \ref{Poyting time evolution}, each point of the field line moves over time with the velocity $\textbf{p}$. Therefore, the system of the equations on the field lines takes the following form:
\begin{align}
    \frac{d \textbf{x}(s,t)}{ds} &= \textbf{e}(\textbf{x}(s,t), t) \label{dx/dt = e} \\
    \frac{d \textbf{x}(s,t)}{dt} &= \textbf{p}(\textbf{x}(s,t), t) \label{dx/dt = p}
\end{align}
Using the equation (\ref{the evolution without divP}), we obtain the following property:
\begin{equation}
    \boxed{\frac{d \textbf{e}(\textbf{x}(s,t),t)}{dt} = \frac{d \textbf{p}(\textbf{x}(s,t),t)}{ds}}
\end{equation}
This property ensures commutativity of flows along $s$ and $t$ in the chosen parametrization.

\section{Knot invariants as invariants of motion \label{knotinvs}}

There are solutions of Maxwell equations that exhibit topologically nontrivial structure of the electric field lines that makes them worth studying. Firstly, some solutions have the property that all/some of the field lines are closed loops - limit cycles. Secondly, a particular closed field line can be topologically nontrivial i.e. it can be a knot. Thirdly, a pair of closed field lines can be linked. 
\\
\\
A closed field line could be identified with some string or strand. Each element of the string moves with velocity equal to the Poynting vector at the corresponding point of the space. As soon as one identifies the field lines with moving strands, natural questions appear. One can ask if the topological structure of the strands is preserved over time. Namely, if a strand was knotted, can it become a different knot? If a pair of strands was linked together, can they become unlinked at a further moment of time? These questions are quite similar and refer to the situation that involves crossings of strands.
\\
\\
There are tools to address this type of questions. They are called knot and link invariants. These invariants are preserved under smooth deformations of knots and links that do not admit crossings of lines. In our work we use the Gauss linking integral \cite{Gauss} which is an example of a link invariant:
\begin{equation}
    \text{Link}_{\textbf{x}, \textbf{y}} = \oint  d x_{\alpha} \oint  d y_{\beta} \, \epsilon_{\alpha \beta \gamma}
   \, \frac{x_{\gamma} - y_{\gamma}}{|\textbf{x} - \textbf{y}|^3}
\end{equation}
We show that the  linking integral computed for a pair of electric field lines \textit{does not} change over time, i.e. it is an integral of the Poynting evolution. This fact gives the answer for the second question, while the answer for the first one involves non-Abelian generalization of the Gauss integral and we leave it for future studies.
\subsection{Gauss linking integral}
\label{Gauss linking integral}
In this section we consider an application of the Gauss linking integral to the study of the evolution of the field lines. We show that the linking integral applied for a pair of limit cycles is preserved under the time evolution of the field lines induces by the Poynting vector.
\\
The linking integral is a topological invariant and it is defined for a pair of closed lines $\textbf{x}(s_x), \textbf{y}(s_y)$:
\begin{equation}
    \label{linking integral}
    \text{Link}_{\textbf{x}, \textbf{y}} = \int ds_x \, ds_y \, \epsilon_{\alpha \beta \gamma} \, 
    \frac{d x_{\alpha}}{ds_x} \, \frac{d y_{\beta}}{ds_y} \, \frac{x_{\gamma} - y_{\gamma}}{|\textbf{x} - \textbf{y}|^3}
\end{equation}
where $s_x, s_y$ are parametrizations and their indices $x,y$ reflect belonging to the particular line. The linking integral computes a numerical invariant widely known as the linking number. Therefore, up to the normalization factor this integral is integer-valued. The linking number shows how many times one line winds around another. \\
We consider a null-field solution of the Maxwell equations. As was shown in s.\ref{peculiarity}, time independent parametrization of the electric field lines is obtained by rescaling the fields as (\ref{e = E/E^2}):
\begin{equation}
     \textbf{e} = \textbf{E} / \textbf{E}^2  
\end{equation}
In this parametrization a field line obeys the following equations:
\begin{align}
    \frac{d \textbf{x}(s,t)}{ds} &= \textbf{e}(\textbf{x}(s,t), t) \label{dx/dt = e} \\
    \frac{d \textbf{x}(s,t)}{dt} &= \textbf{p}(\textbf{x}(s,t), t) \label{dx/dt = p}
\end{align}
The crucial property of this parametrization is that the flows along $s$ and $t$ commute:
\begin{equation}
\label{commut of flows}
    \boxed{\frac{d \textbf{e}(\textbf{x}(s,t),t)}{dt} = \frac{d \textbf{p}(\textbf{x}(s,t),t)}{ds}}
\end{equation}
We pick up two distinct limit cycles $\textbf{x}(s_x,t), \textbf{y}(s_y,t)$ and consider the corresponding linking integral. Taking into account the fact that the tangent vectors to the field lines are rescaled electric fields (\ref{dx/dt = e}), we represent the linking integral in the form:
\begin{equation}
    \label{linking integral with fields}
    \text{Link}_{\textbf{x}(t), \textbf{y}(t)} = - \int ds_x \, ds_y \, \epsilon_{\alpha \beta \gamma} \, 
    e_{\alpha}^{x} \, e_{\beta}^{y} \, \frac{\partial }{\partial x_{\gamma}} \, \frac{1}{|\textbf{x} - \textbf{y}|}
\end{equation}
where we simplify the notation $e_{\alpha}^{x} := e_{\alpha}(\textbf{x}(s_x,t),t)$ and $x_{\gamma} := x_{\gamma}(s_x, t)$. The time dependence of the whole linking integral is encoded in the time dependence of the field lines. Time evolution of the field lines in turn is induced by the Poynting vector (\ref{dx/dt = p}). We also used the fact that the ratio in the integrand (\ref{linking integral}) is a total derivative:
\begin{equation}
    \frac{x_{\gamma} - y_{\gamma}}{|\textbf{x} - \textbf{y}|^3} = -\frac{\partial }{\partial x_{\gamma}} \, \frac{1}{|\textbf{x} - \textbf{y}|}
\end{equation}
To show that the integral $(\ref{linking integral with fields})$ is time independent, we calculate its time derivative:
\begin{equation}
    \frac{d }{dt} \text{Link}_{\textbf{x}(t), \textbf{y}(t)} = \int \, ds_x \, ds_y \, \epsilon_{\alpha \beta \gamma} \left[ 
    \frac{de^{x}_{\alpha}}{dt}  \, e^{y}_{\beta} \, \frac{\partial }{\partial x_{\gamma}} \, \frac{1}{|\textbf{x} - \textbf{y}|} + 
    e^{x}_{\alpha} \,\frac{de^{y}_{\beta}}{dt}  \, \frac{\partial }{\partial x_{\gamma}} \, \frac{1}{|\textbf{x} - \textbf{y}|}
    +
    e^{x}_{\alpha} \, e^{y}_{\beta} \, \frac{d}{dt} \frac{\partial }{\partial x_{\gamma}} \, \frac{1}{|\textbf{x} - \textbf{y}|}
    \right]
\end{equation}
For the first two terms in the integrand we change the derivatives of time to the derivatives along the field line using the property (\ref{commut of flows}). Integrating these terms by parts, we obtain the following expression:
\begin{equation}
    \frac{d }{dt} \text{Link}_{\textbf{x}(t), \textbf{y}(t)} =  \int \, ds_x ds_y \, \epsilon_{\alpha \beta \gamma} \, \left[ e^{y}_{\beta} \Bigl( p^{x}_{\lambda} e^{x}_{\alpha} - p^{x}_{\alpha} e^{x}_{\lambda} \Bigr) - e^{x}_{\alpha} \left( p^{y}_{\lambda} e^{y}_{\beta} - p^{y}_{\beta} e^{y}_{\lambda} \right)
    \right]
    \frac{\partial }{\partial x_{\gamma}} \frac{\partial }{\partial x_{\lambda}} \, \frac{1}{|\textbf{x} - \textbf{y}|} 
\end{equation}
Using the properties of the null-field solutions (\ref{EP - PE}), we have an expression of the form:
\begin{equation}
    \frac{d }{dt} \text{Link}_{\textbf{x}(t), \textbf{y}(t)} = \int \, ds_x ds_y 
    \left[
    \Bigl( e^x_{\alpha} B^y_{\alpha} + e^y_{\alpha} B^x_{\alpha} \Bigr)
    \frac{\partial }{\partial x_{\lambda}} \frac{\partial }{\partial x_{\lambda}} \, \frac{1}{|\textbf{x} - \textbf{y}|}
    -  \left( e_{\alpha}^y B_{\beta}^x + e_{\alpha}^x B_{\beta}^y \right)
    \frac{\partial }{\partial x_{\alpha}} \frac{\partial }{\partial x_{\beta}} \, \frac{1}{|\textbf{x} - \textbf{y}|} 
    \right]
\end{equation}
In the first terms 3d Dirac delta function appears:
\begin{equation}
    \frac{\partial }{\partial x_{\lambda}} \frac{\partial }{\partial x_{\lambda}} \, \frac{1}{|\textbf{x} - \textbf{y}|} = - 4 \pi \delta^{(3)}(\textbf{x} - \textbf{y})
\end{equation}
The last terms turn out to be total derivatives and finally we obtain the following:
\begin{equation}
\begin{split}
     \frac{d }{dt} \text{Link}_{\textbf{x}(t), \textbf{y}(t)} &=  -\int \, ds_x ds_y \
    4 \pi \delta^{(3)}(\textbf{x} - \textbf{y}) \Bigl( (\textbf{e}^x \cdot \textbf{B}^y) + (\textbf{e}^y \cdot \textbf{B}^x ) \Bigr) \\
    & - \int \, ds_x ds_y 
    \left[
    \frac{d}{ds_y}(\textbf{B}^x \cdot \nabla^{x}) \, \frac{1}{|\textbf{x} - \textbf{y}|}
    +
    \frac{d}{ds_x}(\textbf{B}^y \cdot \nabla^{y}) \, \frac{1}{|\textbf{x} - \textbf{y}|}
    \right]
\end{split}
\end{equation}
The first integral may contribute because of the presence of the delta function, but it is multiplied by $\textbf{e}^x \cdot \textbf{B}^y$ and $\textbf{e}^y \cdot \textbf{B}^x$  vanishing when $\textbf{x}-\textbf{y} = 0$ due to the null-field condition (\ref{null field condition}), so the integrand is zero. The second integral is zero because the integrand is a sum of total derivatives and the integral is along the closed line. So we conclude that the linking integral of a pair of electric limit cycles is preserved under the time evolution:
\begin{equation}
     \boxed{\frac{d }{dt} \text{Link}_{\textbf{x}(t), \textbf{y}(t)} = 0}
\end{equation}
We note that the conservation of the linking integral is a consequence of the null-field condition and closeness of the lines. Our reasoning does not involve a direct verification that the field lines do not cross.
\subsection{Non-Abelian knot/link invariants}
At the present moment  the colored HOMFLY-PT polynomial is one of the most powerful knot and link invariants \cite{Alexander, Conway, Jones}. See \cite{Bishler, Bishler_2021} for recent results in this area. It is defined as the vacuum expectation value of the Wilson loop operators along $n$ components of the link $\mathcal{L}$ in the 3d Chern-Simons theory on $S^3$ with the gauge group $SU(N)$:
\begin{equation}
     \label{WilsonLoopExpValue}
         H_{R_1, \ldots, R_n}^{\mathcal{L}_1, \ldots, \mathcal{L}_n} = \left\langle \prod_{i = 1}^{n} \text{tr}_{R_i} \ P \exp \left( \oint_{\mathcal{L}_i} A \right) \right\rangle_{CS},
\end{equation}
where Chern-Simons action is given by
\begin{equation}
    S_{CS}[A] = \frac{k}{4 \pi} \int_{S^3} \text{tr} \left(  A \wedge dA + \frac{2}{3} A \wedge A \wedge A \right).
\end{equation}
Here $R_i$ are the representations of the gauge group and $\mathcal{L}_i$ are link components. \\
The Gauss linking integral naturally appears in the Abelian version of Chern-Simons theory. In the non-Abelian case it also arises in the perturbative expansion of the HOMFLY-PT invariant for a link in the limit $k \rightarrow \infty$ \cite{GMM}. Actually, the non-Abelian generalization provides an infinite family of the knot/link invariants appearing in the perturbative expansion of the HOMFLY-PT invariant. These invariants are widely known as Vassiliev invariants \cite{Vassiliev, ChmutovDuzhin, VasInv} and can be represented as the complicated contour integrals along knots/links. For example, we provide an explicit form of the second Vassiliev invariant of a link component $\mathcal{L}_i$:

\begin{align}
    \begin{aligned}
    \rho(\mathcal{L}_i) &= \frac{1}{8 \pi^2} \oint_{\mathcal{L}_i} dx_{\mu} \int\limits^x dy_{\nu} \int\limits^y dz_{\rho} \int\limits^z dw_{\sigma} \, \epsilon_{\sigma \nu \alpha} \, \epsilon_{\rho \mu \beta} \, \frac{(w - y)_{\alpha} (z - x)_{\beta}}{|\textbf{w} - \textbf{y}|^3 |\textbf{z} - \textbf{x}|^3} \\
    &-\frac{1}{32 \pi^3} \oint_{\mathcal{L}_i} dx_{\mu} \int\limits^x dy_{\nu} \int\limits^y dz_{\rho} \, \epsilon_{\alpha \beta \gamma} \,
    \epsilon_{\mu \alpha \sigma} \, \epsilon_{\nu \beta \lambda} \, \epsilon_{\rho \gamma \tau} \int\limits_{\mathbb{R}^3} d^3w
    \frac{(w - x)_{\alpha} (w - y)_{\lambda} (w - z)_{\tau}}{|\textbf{w} - \textbf{x}|^3 |\textbf{w} - \textbf{y}|^3 |\textbf{w} - \textbf{z}|^3}
    \end{aligned}
\end{align}
The similar analysis as in the section \ref{Gauss linking integral} can be carried out for these integrals when the curve $\mathcal{L}_i$ is the field line.  The application of this invariant will give answers to the questions about knot structure of the evolving field lines. We leave this question for future studies.
\\
These integral-type invariants are obtained in the Lorentz gauge. It would be interesting to make a connection with other gauges and lift-up the story to the level of the $\mathcal{R}$-matrices. For possible approach see \cite{Smirnov}.

\section{Integrability \label{integrability}}
It is a well known fact that the Gauss linking integral is connected to the electric/magnetic helicity \cite{Arrayas}:
\begin{equation}
    h_e = \int_{\mathbb{R}^3} d\textbf{x} \int_{\mathbb{R}^3} d\textbf{y} \ \textbf{E}(\textbf{x}) \cdot \frac{\textbf{E}(\textbf{y}) \times (\textbf{x} -\textbf{y})}{|\textbf{x} -\textbf{y}|^3}
\end{equation}
\begin{equation}
    h_m = \int_{\mathbb{R}^3} d\textbf{x} \int_{\mathbb{R}^3} d\textbf{y} \ \textbf{B}(\textbf{x}) \cdot \frac{\textbf{B}(\textbf{y}) \times (\textbf{x} -\textbf{y})}{|\textbf{x} -\textbf{y}|^3}
\end{equation}
These formulas are similar to the Gauss linking integral for two closed field lines (\ref{linking integral with fields}). In this terms the helicity can be obtained by "summing" the linking integrals over all pairs of field lines:
\begin{equation}
    h_e = \sum_{i \not =  j} \ 
    \oint ds_i \oint ds_j \ \textbf{E}(\textbf{x}_i) \cdot \frac{\textbf{E}(\textbf{x}_j) \times (\textbf{x}_i -\textbf{x}_j)}{|\textbf{x}_i -\textbf{x}_j|^3}
\end{equation}
Beside the fact that the sum over field lines is not rigorously defined, one can think of the helicity as the average linking number of field lines.
It turns out that these quantities are integrals of motion in the null-field solutions of Maxwell equations:
\begin{equation}
    \frac{dh_{e}}{dt} = \frac{dh_{m}}{dt} = \int_{\mathbb{R}^3} d\textbf{x} \ \textbf{E}(\textbf{x}) \cdot \textbf{B}(\textbf{x}) = 0
\end{equation}
If we believe in conservation on knot/link invariants for closed field lines, then the invariants of knots/links should be connected with nontrivial integrals of motion. It would imply that the system of Maxwell equations coupled with null-field condition is actually integrable.

\section{Electromagnetic knots}
In this section we consider particular examples of solutions of the Maxwell equations that exhibit nontrivial behaviour of the field lines. In recent papers \cite{Lechtenfeld:2017tif, Kumar:2020xjr} a family of source-free finite action solutions of Maxwell equations was constructed. The solutions are interesting because the field lines are similar to knots and links. The electric and magnetic fields as functions of the coordinates are the rational functions and therefore the solutions are called rational electromagnetic fields. We note that the zeros of the denominator are complex, so the solutions do not have singularities.
\subsection{Ranada's Hopf solution \label{Hopfsol}}
\begin{figure}[h!]
    \centering
    \includegraphics[scale=0.255]{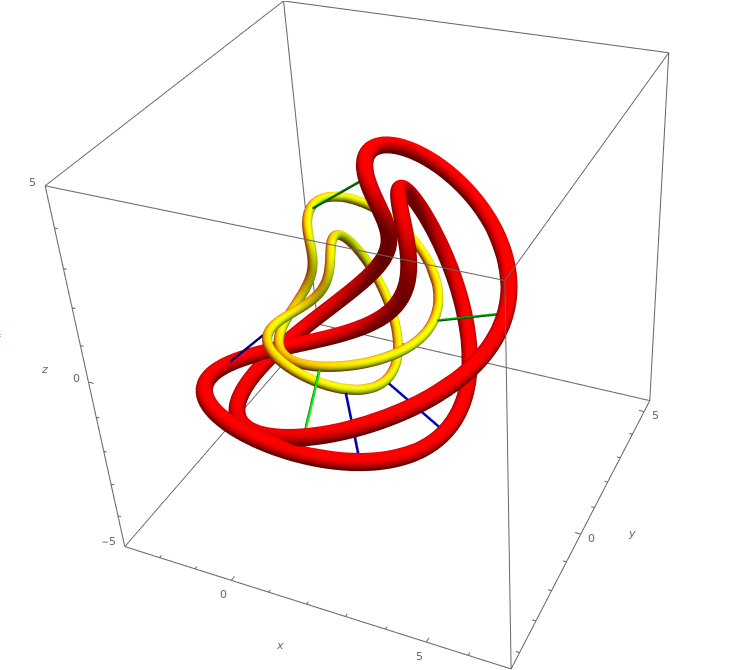}
    \includegraphics[scale=0.255]{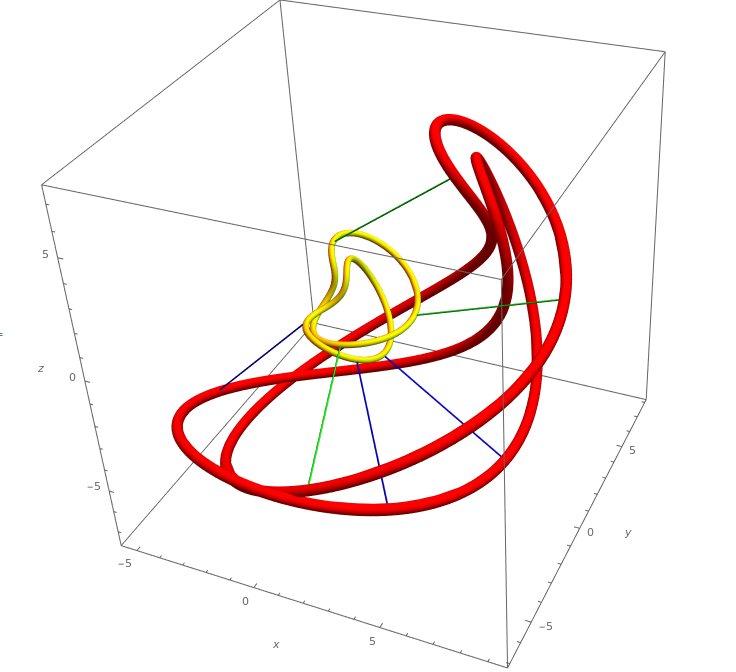}
    \caption{\footnotesize The yellow lines evolve into red lines over time. There are pictures of two successive moments of time. The green and the blue curves are the trajectories. The link is preserved over time. }
    \label{field lines hopf link}
\end{figure}

Our main example is the celebrated Hopf-Ranada solution:
\begin{equation}
\label{Hopf}
    \textbf{E} + \mathrm{i} \textbf{B} =\frac{1}{\left((t-\mathrm{i})^{2}-r^{2}\right)^{3}}\left(\begin{array}{c}
(x-\mathrm{i} y)^{2}-(t-\mathrm{i}-z)^{2} \\
\mathrm{i}(x-\mathrm{i} y)^{2}+\mathrm{i}(t-\mathrm{i}-z)^{2} \\
-2(x-\mathrm{i} y)(t-\mathrm{i}-z)
\end{array}\right)
\end{equation}
This solution is the first member of the family. Here we list the properties of the field lines that were observed on the computer simulations (see Fig.\ref{field lines hopfion}):
\begin{figure}[h!]
    \centering
    \includegraphics[scale=0.3]{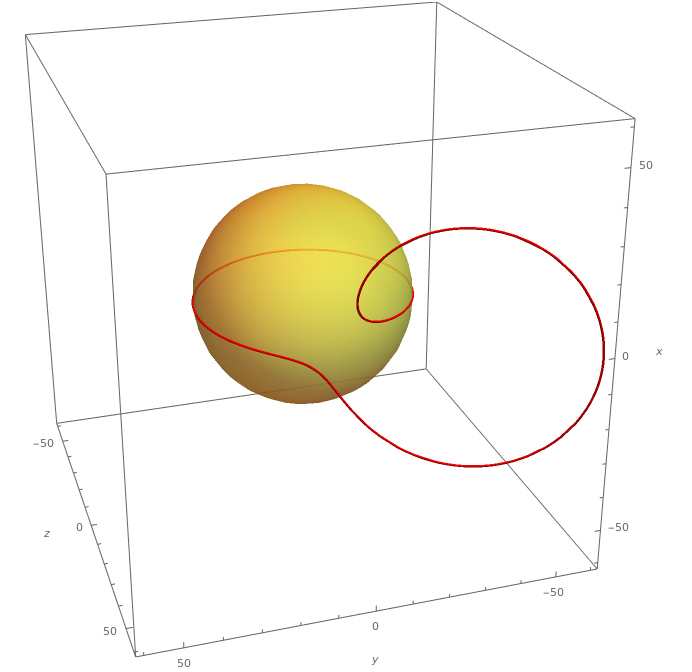}
    \includegraphics[scale=0.3]{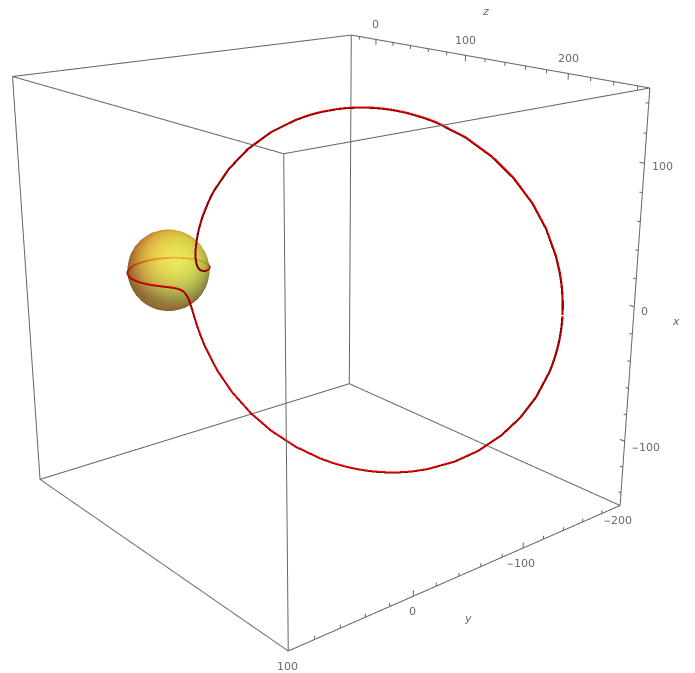}
    \caption{\footnotesize The yellow sphere is the light cone $x^2 + y^2 + z^2 = t^2$. The red lines are electric field lines at the moment $t = 30$. A part of the field line lies on the equator of the sphere. The other part tends to form a circle.}
    \label{field lines hopfion}
\end{figure}
\begin{itemize}
    \item All field lines are closed.
    \item Every field line is topologically unknot.
    \item Any two field lines are linked.
    \item Links are preserved over time (see Fig.\ref{field lines hopf link})
\end{itemize}
The last observation was a hint to consider the Gauss linking integral and demonstrate this property analytically. To make an attempt to explain the other properties, we consider the simplified version of the electric field, namely the electric field far from the origin. On the coordinate scales much larger that the time $t$ the electric field has the form:
\begin{equation}
\label{large scales hopfion}
\textbf{E}_{\infty}(\textbf{x}) = \frac{1}{r^6}\left(\begin{array}{c}
    -x^{2} + y^2 + z^{2} \\
    -2 x y \\    
    -2 x z
\end{array}\right)
\end{equation}
This electric field defines an integrable system of differential equations on the field lines and the solution  has the following form:
\begin{equation}
    \label{circle field lines at large scales hopfion}
    x^2 + (y \cos \theta + z \sin \theta - b)^2 = b^2
\end{equation}
where all values of the radius $b$ and the angle $\theta$ are possible. The circles are normal to the plane $yz$, pass through the origin and their centers lie on the plane $yz$. This fact is in agreement with the Fig.\ref{field lines hopfion} because one can see that a part of a field line is similar to a circle (\ref{circle field lines at large scales hopfion}). 
\\
\\
We note that the simplified version of the electric field at large scales (\ref{large scales hopfion}) defines closed field lines (\ref{circle field lines at large scales hopfion}).
The fact that the simplified field at large scales defines an integrable system of differential equations and closed field lines simultaneously might not be by chance. The integrability is connected with existence of the function that is constant along field lines (\ref{circle field lines at large scales hopfion}):
\begin{equation}
    h(\textbf{x}) = \frac{\sqrt{y^2 + z^2}}{x^2 + y^2 + z^2}
\end{equation}
\begin{equation}
    \frac{d}{ds} h(\textbf{x}(s)) = 0, \hspace{10mm} \text{where} \hspace{10mm} \frac{d \textbf{x}(s)}{ds} = \textbf{E}_{\infty}(\textbf{x})
\end{equation}
The simplified version of the field (\ref{large scales hopfion}) is effectively two dimensional and one "integral of motion" $h(\textbf{x})$ is sufficient to determine the field lines. In case of the full Hopf-Ranada solution two "integrals" might explain the closedness of the field lines.

\subsection{Other knots/links \label{othersols}}

\begin{figure}[h!]
    \centering
    \includegraphics[scale=0.22]{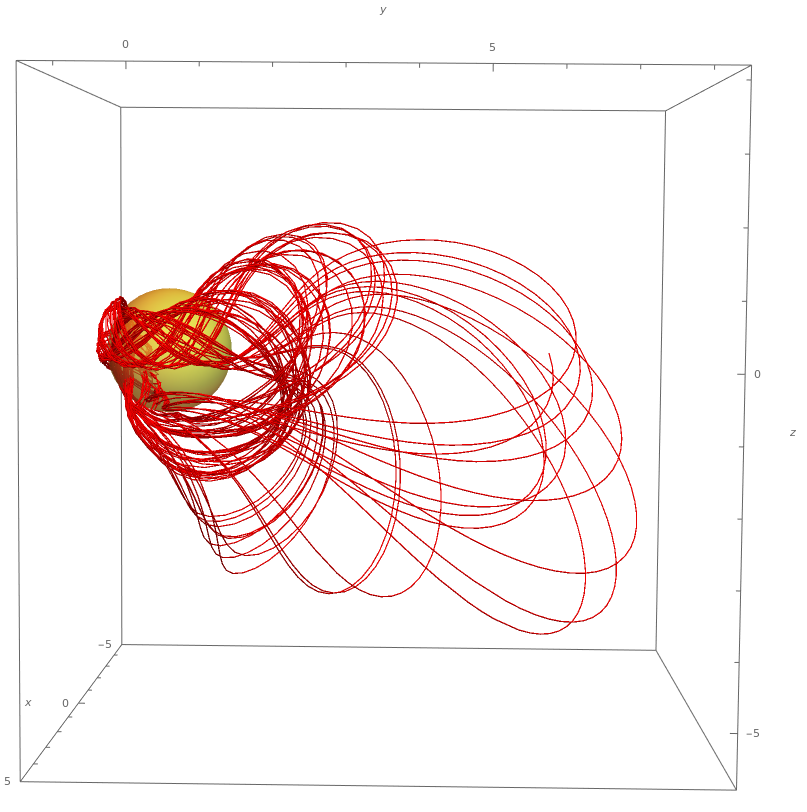}
    \includegraphics[scale=0.22]{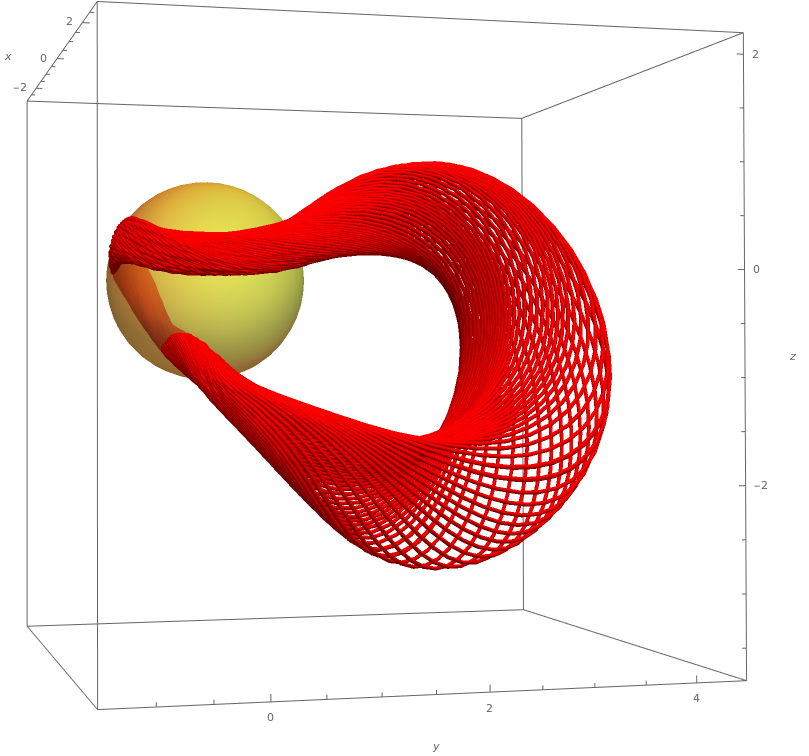}
    \caption{\footnotesize The left and right pictures correspond to the same solution of the Maxwell equation. We conjecture that almost all field lines have infinite length. They can be tightly wound as in the left picture or have a form of an attractor as in the right picture.}
    \label{field lines higher solution of general position}
\end{figure}

\begin{figure}[h!]
    \centering
    \includegraphics[scale=0.22]{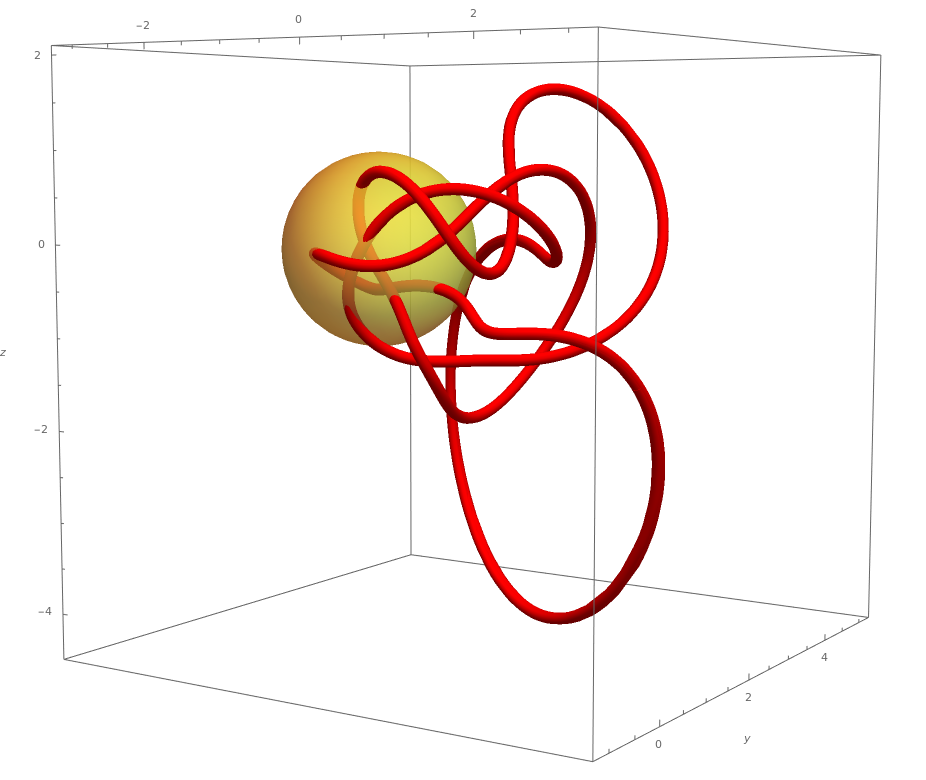}
    \includegraphics[scale=0.23]{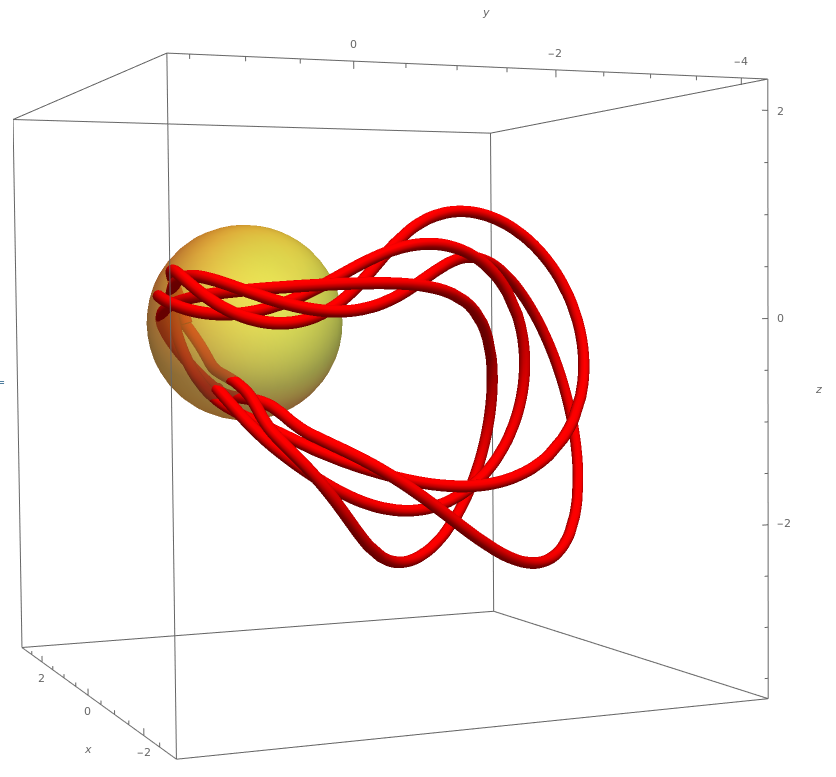}
    \caption{\footnotesize These pictures correspond to the exceptional closed field lines. The left and the right field lines correspond to the same solution of the Maxwell equation. This solution is a higher member of the family. One can see that the previous pictures Fig.\ref{field lines higher solution of general position} represent non-closed field lines that wind around these closed ones. The yellow sphere is the light cone $x^2 + y^2 + z^2 = t^2$.}
    \label{field lines higher solution}
\end{figure}

The higher members of the family of knot solutions exhibit more complicated structure of the field lines. There are rare closed limit cycles in space and they represent various knots (see Fig.\ref{field lines higher solution}). Almost all field lines are tightly wound around the limit cycles and we conjecture that they have infinite length (see Fig.\ref{field lines higher solution of general position}).
It would be interesting to find an efficient description of this complicated structure and classify knots and links that appear in null-field solutions. For a possible approach to the problem see \cite{Kumar:2020xjr}.

\section{Conclusion \label{conclusion}}  
Long ago hidden integrability was predicted to be one of the governing principles
for dynamics in stringy models \cite{UFN2, Morozov1995, Mironov:1993,Mironov:1994}.
Since then this was proved to be the case in quite a number of examples.
In this paper we suggest to search for integrability in generic solutions of
Maxwell equations.  
The point is to reformulate Maxwell dynamics in terms of behavior of the
field lines. Behaviour of the field lines was also discussed in \cite{Jehle:1971pu,Jehle:1972zk,Jehle:1974he,Jehle:1976ru}.
At a given time we can define a system of non-intersecting world lines,
and Maxwell dynamics convert them onto a system of world surfaces.
Accordingly there are two directions $s$ and $t$ and two kinds of dynamics  
and, potentially, integrability -- for $s$ and $t$-evolutions.
We provided some evidence that both kinds of integrability are really present,
and $t$-integrability is related to conservation of topological invariants,
like knot polynomials and Vassiliev coefficients of their expansions.
It would be very interesting to develop these arguments into a full-fledged
theory -- most probably this will include non-Abelian considerations.
In particular, at the $SU(2)$ level one can exploit peculiar properties
of Lorentz and conformal groups in $4d$ \cite{Lechtenfeld:2017tif}.
In another direction, one can relate evolution of field lines with
loop equations and integrability of eigenvalue matrix models \cite{UFN3, Morozov:2005, Mironov:2002}.
We hope for a new and profound progress on these issues,
which would enrich our understanding of integrability of effective theories
with the help of a very concrete and down-to-earth story of
Maxwell equations.

\section*{Acknowledgements}
We appreciate clarifying discussions with N.Kolganov, S.Mironov, V.Mishnyakov, And.Morozov, A.Popolitiov, A.Sleptsov and N.Ushakov at early stages of this project. 
\\
Our work was partly supported by the grant of the Foundation for the Advancement of Theoretical Physics “BASIS" (N.T.), by RFBR grants 19-02-00815 (A.M.), 20-01-00644 (N.T.), by joint RFBR grants 19-51-53014-GFEN (A.M.), 19-51-18006-Bolg (A.M.), 21-51-46010-CT (A.M, N.T.).

\printbibliography

\end{document}